\documentclass[smallextended]{svjour3}       

\usepackage{times}

\usepackage{url}
\usepackage[usenames,dvipsnames]{color}
\usepackage{algorithm} 
\usepackage[noend]{algpseudocode}
\usepackage{multirow} 
\usepackage[square,comma,numbers,sort]{natbib}
\usepackage[rm]{caption} 
\DeclareCaptionType{copyrightbox}
\usepackage{booktabs}
\usepackage{balance}
\usepackage{xspace}
\usepackage{amssymb}
\usepackage{graphicx}

\newcommand{\methfont}[1]{{\emph{#1}}}
\newcommand{\typesimple}{\methfont{Constant}}
\newcommand{\typeboolean}{\methfont{Boolean}}
\newcommand{\typeconjunctive}{\methfont{Conjunctive-Ranked}}
\newcommand{\typeranked}{\methfont{Normal-Ranked}}
\newcommand{\typetwolevel}{\methfont{Multi-Level-Ranked}}
\newcommand{\coll}{\mbox{\ensuremath{\cal{D}}}}
\newcommand{\judg}{\mbox{\ensuremath{\cal{J}}}}
\newcommand{\quer}{\mbox{\ensuremath{\cal{Q}}}}

\newcommand{\mn}[1]{\ensuremath{\mathnormal{#1}}}

\newcommand{\metricfont}[1]{{\small\sf{#1}}}
\newcommand{\metric}[1]{\metricfont{#1}}
\newcommand{\RBP}{\metric{RBP}}
\newcommand{\AP}{\metric{AP}}
\newcommand{\NDCG}{\metric{NDCG}}
\newcommand{\ERR}{\metric{ERR}}
\newcommand{\med}[1]{{\ensuremath\metric{MED}_{\small\metricfont{#1}}}}

\def\D{\hphantom{1}}

\newcommand\method[1]{{\sf\small{#1}}}

\newcommand{\indri}{\method{Indri\xspace}}

\newcommand\gb[1]{$#1$\,GB}

\def\D{\hphantom{1}}

\newcommand{\myurl}[1]{{\url{#1}}}
\newcommand{\mycaption}[1]{\caption{\normalfont{#1}}}
\newcommand{\myparagraph}[1]{\paragraph*{#1}~\\\noindent}

\newcommand{\mycomment}[1]{}

\newlength{\onedigit}
\newcommand{\showgraph}[1]{\includegraphics[scale=0.40]{#1}}

\begin{document}

\titlerunning{Efficiency-Effectiveness Tradeoffs in
Multi-Stage Retrieval}
\authorrunning{Clarke, Culpepper, Moffat}

\title{
Assessing Efficiency-Effectiveness Tradeoffs in Multi-Stage Retrieval
Systems Without Using Relevance Judgments
} 

\author{ 
Charles L. A. Clarke \and J. Shane Culpepper \and Alistair Moffat
}

\institute{
Charles L. A. Clarke\\
University of Waterloo, Canada\\
\email{claclarke@cs.uwaterloo.ca}\\
\\
J. Shane Culpepper\\
RMIT University, Australia\\
\email{shane.culpepper@rmit.edu.au}\\
\\
Alistair Moffat \\
The University of Melbourne, Australia\\
\email{ammoffat@unimelb.edu.au}
}

\date{Received: date / Accepted: date}

\maketitle


\begin{abstract}
Large-scale retrieval systems are often implemented as a cascading
sequence of phases -- a first filtering step, in which a large set of
candidate documents are extracted using a simple technique such as
Boolean matching and/or static document scores; and then one or more
ranking steps, in which the pool of documents retrieved by the filter
is scored more precisely using dozens or perhaps hundreds of
different features.
The documents returned to the user are then taken from the head of
the final ranked list.
Here we examine methods for measuring the quality of filtering and
preliminary ranking stages, and show how to use these measurements to
tune the overall performance of the system.
Standard top-weighted metrics used for overall system evaluation are
not appropriate for assessing filtering stages, since the output is a
set of documents, rather than an ordered sequence of documents.
Instead, we use an approach in which a quality score is computed
based on the discrepancy between filtered and full evaluation.
Unlike previous approaches, our methods do not require relevance
judgments, and thus can be used with virtually any query set.
We show that this quality score directly correlates with actual
differences in measured effectiveness when relevance judgments are
available.
Since the quality score does not require relevance judgments, it can
be used to identify queries that perform particularly poorly for a
given filter.
Using these methods, we explore a wide range of filtering options
using thousands of queries, categorize the relative merits of the
different approaches, and identify useful parameter combinations.
\end{abstract}


\section{Introduction}
\label{sec-intro}

The purpose of an information retrieval system is well-defined: given
a query $q$, and a large collection $\coll$ of documents,
identify and present a small subset of the collection by
identifying documents that are deemed responsive to $q$.
Typical collections contain billions of documents and occupy
terabytes of storage.
Queries are a often a few words long, and answer lists contain between 
ten and a few hundred items.
{\citet{zm06compsurv}} and {\citet{bcc10ir}}
provide surveys of these processes.

The systems that have been developed in response to these goals are
assembled from a suite of possible components, with many different
ways of generating query-document similarity scores so that documents
can be ranked.
But in broad terms, most retrieval systems can be thought of as being
composed from a small number of underlying building blocks that can
be categorized into three groups: {\emph{pre-ordering stages}}, that
make use of static index-time data such as page-rank or document
quality scores (including spam scores); {\emph{filtering stages}}, in
which a subset of the collection is extracted; and then one or more
{\emph{detailed ranking phases}}, in which a comprehensive evaluation
of those documents is undertaken, so that the final top-$k$ ranked
list can be generated.
Each of these stages can in turn be composed of one or more steps.
In this work we differentiate only between the filtering stage that
identifies potential candidate documents, and the final ranking stage
that works with a smaller subset of documents drawn from the entire
collection.

The end-to-end effectiveness of an IR system is measured
using any one of a large range of metrics such as precision (\metric{P}),
average precision (\AP),
discounted cumulative gain (\metric{DCG}),
normalized discounted cumulative gain (\NDCG),
rank-biased precision (\RBP),
the Q-Measure (\metric{QM}),
reciprocal rank (\metric{RR}),
expected reciprocal rank (\ERR),
time-biased gain (\metric{TBG}),
and so on {\citep{jk02acmtois,mz08acmtois,sk08irj,cmzg09cikm,sc12sigir}}.
Two broad families can be identified -- metrics that are
{\emph{recall dependent}}, for which the calculated score is relative
to the best that could be attained by any ranking, and hence can be
regarded as an absolute assessment ({\AP}, {\NDCG},
{\metric{QM}}); and metrics that are {\emph{recall independent}}, for
which the calculated score reflects the user's experience of the supplied
ranking, and is not affected by the density or otherwise of relevant
documents in the unseen part of the collection ({\metric{P}}, {\metric{DCG}},
{\RBP}, {\metric{RR}}, {\ERR}, {\metric{TBG}}).

Conventional metrics are not directly applicable to the task of
measuring the quality (that is, contribution to overall search
performance) of the pre-ordering stage, or of the filtering stage.
For example, Boolean conjunction has been suggested as a useful
filtering stage, so that every presented answer document has every
query term in it somewhere {\citep{kt14sigir}}.
But applying {\AP} or {\NDCG} to the output of a Boolean conjunction
is unhelpful, since before it is seen by the user, that set of
documents might be (and quite probably will be) permuted in to an
entirely different ordering by the ranking stages.

\myparagraph{Contributions}
We make two key contributions.
First, we demonstrate the applicability of a filter-stage evaluation
approach that is based on bounding the loss of end-to-end
effectiveness that can occur as a result of the filtering process.
This quantification of the possible degradation is straightforward to
compute for all recall-independent metrics; and, in the experiments
described below, is validated using data and topics for which
relevance judgments are available.
Second, we then use that approach to examine thousands of queries
for which relevance judgments are not available, and
compare a variety of possible early-stage filters.

Those results then provide the basis for estimating the combined
multi-stage retrieval time versus effectiveness trade-off options for
large-scale retrieval.
Extending prior results~\cite{al13irj}, our findings show that
conjunctive Boolean mechanisms are unlikely to provide useful
trade-offs, despite their high retrieval speed.
On the other hand, aggressive WAND evaluation strategies, in which
answer list ``safeness'' is sacrificed for evaluation speed, do offer
appealing options.

\section{Background}

A retrieval system can be regarded as being a composition of multiple
phases, with any of the phases potentially being omitted, or broken
in to sub-phases.
We now group possible retrieval tasks into the three general phases;
then describe each phase and the systems that result when they are
composed in various ways; and finally in this section discuss ways in
which system effectiveness can be evaluated.

\myparagraph{Static Ordering}
The first phase is a method for {\emph{static ranking}}, or
{\emph{pre-ordering}} the documents in the collection.
A function $P(\coll,k)$ is provided that processes $\coll$ according
to some query-independent scoring regime, and then returns as an
ordered set the $k$ documents with the highest static scores.
The abbreviated notation $P(\coll)$ indicates $P(\coll,|\coll|)$, a
complete ordering of $\coll$ according to static score, rather than a
top-$k$ truncated ranking.
Examples of this first phase include orderings based on page rank,
spam score, document length, perceived document quality, plus
combinations of these and other query-agnostic features.

\myparagraph{Filtering}
The second phase of a retrieval system is a {\emph{filtering}} stage
$F(\coll,q,k)$ that extracts a $k$-subset of the collection $\coll$
in response to a query $q$, usually (but not always) selecting those
that match $q$ in some manner.
At most $k$ documents are selected by the filter, which retains the
same ordering of documents in the output as was supplied in the
input.
Again, as a notational convenience, we use $F(\coll,q)$ for
$F(\coll,q,|\coll|)$ to cover situations in which all of the
documents in $\coll$ might potentially be allowed through by the
filter.
A trivial filter could simply transmit the first $k$ documents that
are presented, regardless of their merits, and withhold the remainder
of the collection.
A more complex filter might implement a conjunctive {\metric{AND}}
operation that selects the set of documents that contain all of the
query terms.
Or third, a simple-but-fast ranking system might be employed as a
filter, passing through the top (say) $k=10^3$ documents according to
a scoring mechanism such as term overlap count.

\myparagraph{Ranking}
The third phase of a retrieval system is a {\emph{ranking}}
or {\emph{reordering}} stage $R(\coll,q,k)$ that selects at most 
$k$ documents from $\coll$,
according to their perceived relevance to query $q$, and returns as
an ordered sequence the top $k$ documents in decreasing score order.
For example, a BM25 or Language Model similarity computation based on
collection, term, and document statistics might be used to score each
document in the collection (see {\citet{zm06compsurv}} or
{\citet[Chapters~8 and~9]{bcc10ir}}), and then the documents
with the top $k$ scores extracted and returned.
As before, the shorthand $R(\coll,q)$ is used to represent
$R(\coll,q,|\coll|)$, with (potentially) every document in the
collection assigned a score and then returned by the ranker.


\myparagraph{System Options}
These different processes might be implemented using different
underlying structures and mechanisms.
For example, the pre-scoring process might happen at index
construction time, and involve explicit reordering of the documents
in the collection so that high-quality documents have low document
numbers.
The filtering stage might then be supported by a document-level
(non-positional) inverted index, so that documents matching the
filter's specification can be quickly identified.
The ranking stage might be supported by the same index augmented by
term positions; or, especially if complex features based on phrases
or term proximities are being employed, might be based on document
surrogates computed at the time the collection was indexed (a
``direct'' file).

\newcommand{\labelledpar}{\smallskip\par\noindent---~}
Combining the various retrieval options in different ways gives a range
of possible retrieval systems.

\labelledpar\typesimple:
A simple ordered list, such as a newspaper home page.
In this configuration, $\coll$ is the set of newspaper
stories ordered temporally; $P()$ is a query-free
scoring regime based on story recency and story importance as
determined by non-query features, such as user profile or
past reading behavior; and the retrieval system is of the
form $P(\coll,k)$, where $k\approx10$, for example, is the number of
stories to be presented.

\labelledpar\typeboolean:
General Boolean querying, such as 1980s-style abstract search
systems.
In this configuration, $\coll$ is the set of abstracts; $q$
is expressed as a conjunction of concepts, each described
as a disjunction of terms or negated terms; $F()$ is a Boolean
matching process; and the retrieval system is of the form
$F(\coll,q)$; in this configuration there is no limit on the 
number of answers provided, and no particular ordering within 
the documents that match the query.

\labelledpar\typeconjunctive:
Ordered conjunctive querying, of a style used by some early
search systems.
In this configuration, $\coll$ is the set of web pages; $q$
is a list of query terms; $F()$ is a Boolean conjunction over
those terms; and the retrieval system is of the form
$F(P(\coll),q,k)$, where $k\approx10$ and $P()$ provides a
static collection-wide ordering of some sort.

\labelledpar\typeranked:
Standard document ranking, such as the systems developed by academic
groups participating in TREC Ad-Hoc and Web Tracks.
In this configuration, $\coll$ is the set of web pages; $q$ is a list
of query terms; $R()$ is a ranking computation; and the retrieval
system has the form $R(\coll,q,k)$, where $k$ is of the order of
$1{,}000$.

\labelledpar\typetwolevel:
Multi-phase systems, including the ones presumed to be deployed by
commercial web search providers.
The retrieval system has the form $R(F(\coll ,q , k_1), q, k_2)$,
where $k_1\approx 10^3$, and $k_1\gg k_2\approx 10$ or $20$.
The filter $F()$ might be a Boolean conjunction {\citep{kt14sigir}},
or might be a simple-but-fast ranker {\citep{wf14cikm}} that returns
not more than $k_1$ (and at least $k_2$ documents) from the
collection.
The ranker $R()$ then selects and returns the top $k_2$ of those
$k_1$ documents, employing a large set of features, possibly
including term proximities and adjacencies, and possibly including
the application of learning-to-rank techniques.
There is no requirement that the top $k_2$ documents are a proper
subset of the $k_1$ documents returned from any single filter.
It is entirely possible for the candidate document set $k_1$ to be a
composition of documents drawn from several different filtering
approaches.

\medskip\par\noindent
Note that all of these descriptions are declarative rather than
procedural, and define the result of the computation, not a literal
prescription as to how the computation is to be implemented.
In an actual implementation the phases' computations might be tightly
integrated into one process; or they might be separated.
What is of interest is the logical structure of the computation, not
the implementation details.

\myparagraph{Effectiveness Evaluation}
An {\emph{effectiveness metric}} is a function
$\metric{M}(\coll_q,\judg_q)\rightarrow v$, where $\coll_q$ is an
ordered or unordered sub-collection generated by some ranking system
for query $q$; $\judg_q\subseteq\coll$ is a set of corresponding
positive judgments for query $q$; and $v\in{\cal{R}}$ is a
real-valued number, usually in the range zero to one.
To compute a numeric effectiveness score, each element in $\coll_q$
is checked against~$\judg_q$.
Suppose that $d_i$ is the $i$\,th element of $\coll_q$, whether
$\coll_q$ is ordered or unordered.
Then the (binary) relevance at the $i$\,th position of query $q$'s
ranking is given by
\[
	r_i = \left\{
		\begin{array}{ccl}
		1 && \mbox{if $d_i \in \judg_q$}\\
		0 && \mbox{otherwise\,.}
		\end{array}
		\right.
\]
Different metrics can be defined in terms of which $r_i$ values are
considered, and how they are combined.
Relevance score $r_i$ can also be regarded as being fractional
rather than integral, with $0\le r_i\le1$ indicating the strength of
relevance of $d_i$, a situation known as {\emph{graded relevance}}.
There are also evaluation situations in which $r_i$ is dependent not
just on $d_i$, but on $d_1$ through to $d_{i-1}$ as well, considered
as a set.
For example, the first time a particular document is encountered in a
ranking, or a particular intent interpretation of a query is
encountered, the relevance score $r_i$ might be $1$.
But a subsequent listing of a duplicate document, or of a document
that addresses the same intent, might be considered as corresponding to
$r_i=0$.
One way of interpreting the value $r_i$ is that it is the
{\emph{utility}}, or {\emph{benefit}} that the user accrues by
encountering that document in the ranking.
In this work we assume that the $r_i$ values can be computed in some
appropriate manner, taking into account the supplied ranking, and
that what is required is that the set of $r_i$ values be converted to
a numeric effectiveness score.

\myparagraph{Metrics}
A large number of effectiveness metrics have been proposed.
{\metric{Precision}} is a standard metric applied to {\typeboolean}
querying (which has as its output a set rather than a sequence),
defined as the fraction of $\coll_q$ that appears in $\judg_q$.
Similarly, {\metric{Recall}} is the fraction of $\judg_q$ that
appears in $\coll_q$; and from these two, {\metric{F$_1$}} is
calculated as the harmonic mean of {\metric{Precision}} and
{\metric{Recall}}.
Variants of precision and recall can also be used to provide
effectiveness scores for the ranked sequences generated by the
{\typesimple}, {\typeconjunctive} and {\typeranked} retrieval systems
defined earlier in this section, with scoring typically computed ``to
depth $k$'', for some chosen value of $k$.
Evaluating precision in this way results in a {\emph{top-weighted}}
effectiveness metric, since it is biased in favor of the top-$k$
documents.
Within the $k$ documents, however, all are treated equally.

A range of other top-weighted metrics have been defined, with
behaviors that provide a smoother transition from top to bottom of
the ranking than does $\metric{Precision}@k$.
These include average precision, {\AP}; normalized
discounted cumulative gain, {\NDCG} {\citep{jk02acmtois}};
rank-biased precision, {\RBP} {\citep{mz08acmtois}}; the
Q-measure, {\metric{QM}} {\citep{sk08irj}}; and time-biased gain,
{\metric{TBG}} {\citep{sc12sigir}}.
Each of these top-weighted metrics can be applied to whole rankings,
or, with varying degrees of imprecision (and varying degrees of
knowledge about the magnitude of that imprecision), to fixed-length
depth-$k$ prefixes.

Provided that the required relevance judgments $\judg_q$ are
available to depth $k$ for each member $q$ of a set of test topics
$\quer$, two rankers $R_a()$ and $R_b()$ can be compared by measuring
the relative performance, and calculating a paired significance
test over the set of corresponding values of an effectiveness metric
{\metric{M}}:
\[
	\langle \metric{M}(R_a(\coll,q,k),\judg_q),
	        \metric{M}(R_b(\coll,q,k),\judg_q)
	\mid
		q \in \quer
	\rangle
		\, .
\]

\myparagraph{Judgments}
A critical question that arises is how the judgments $\judg_q$ are
formed.
In most evaluations human judges examine documents, deciding for
each whether or not it is relevant or irrelevant.
Construction of exhaustive judgments is impossibly expensive, and so
only subsets of the collection are normally judged, with
{\emph{pooling to depth $\ell$}} one strategy that can be used to
control the cost of forming judgments so that a set of systems can be
evaluated.
In this approach, a document is judged for a query if and only if its
minimum rank in any of the system runs being pooled is $\ell$ or
less.
Hence, the set $\judg_q$ of judged relevant documents for a query $q$
is likely to be a subset of the true set of relevant documents
{\citep{zob98sigir}}.
As a result, the documents in $\coll_q$ will fall in to one of three
categories: those judged relevant; those judged non-relevant; and
those that are unjudged.
One way of handling unjudged documents is to remove them from the
ranking and form a {\emph{condensed}} sequence; the more common
mechanism is to assume that unjudged documents are non-relevant.
Weighted-precision metrics, including {\RBP}, provide the capacity to
retain a ``residual'' that reflects the range of scores that could
arise, with the size of that range depending on where in the ranking
the unjudged documents appear, and reflecting the maximum and minimum
score that could be achieved if those documents were respectively found
(were they to be judged) to be relevant or not
relevant~{\citep{mz08acmtois}}.

If the evaluation depth $k$ is less than the pooling depth $\ell$,
all documents required during scoring of the contributing runs
will have been judged, and the comparison is
unlikely to be biased.
On the other hand, if the evaluation depth $k$ exceeds the pooling
depth $\ell$, or if runs are to be scored that were not included when
the pool was formed, then comparing effectiveness scores may not be
appropriate without also allowing for the unjudged documents
{\citep{zob98sigir}}.
All three options -- assuming that unjudged implies not relevant;
condensing the sequence; or maintaining error ranges on scores --
have drawbacks.

\myparagraph{Safe Filtering}
In some special cases a filter might be matched to a corresponding
ranker and be said to be {\emph{set safe to depth $k$}}, in that it
selects exactly the final $k$ documents as a set, but not necessarily
in the same final order as assigned by the ranker.
That is, $F()$ and $R()$ are a ``set safe to depth $k$'' combination
if, when considered as sequences,
\[
	R(\coll,q,k) = 
	R(F(\coll,q,k),q) \, .
\]
If this is the case, then the imposition of the filtering stage will not
affect overall retrieval effectiveness {\citep{tf95ipm}}.

But the more usual situation is when both the filter $F()$ and the
ranker $R()$ contribute to the quality of the overall results listings
presented to the users of the system.
Given the large number of ways that filters (which select an
unordered subset of the collection, possibly of bounded size) and
rankers (which select an ordered subset of the collection, again
possibly of bounded size) can be implemented, mechanisms are required
that allow rankers and filters to be compared, and/or their behavior
numerically quantified.

\section{Multi-Phase Effectiveness}

Now consider effectiveness evaluation for a system composed of
distinct filtering and reranking phases.
One possibility is to treat retrieval as an end-to-end process that
creates a ranked sequence; and then interpret the output list
$R(F(\coll,q,k_1),q,k_2)$ exactly the same way as the result of a
{\typeranked} system would be.
That is, if overall performance is all that matters, then assessment
can be via a standard metric $\metric{M}$, and the system can be
regarded as being a single entity, despite the fact that it is
assembled from distinct components.

But suppose that the measurement must focus on the usefulness of a 
filtering stage $F()$, so as to establish (for example) a separate
effectiveness-efficiency trade-off curve for it; or to understand what
effect $k_1$ has on overall effectiveness.
Or, as a second scenario, suppose that alternative filters $F_a()$
and $F_b()$ are provided, and that their usefulness is to be compared
in the context of a specified third-phase ranker.
How should the quality of a pre-ordering, or of a filter, be
measured?

\myparagraph{Recall}
One possible approach is to use available relevance judgments $\judg$
to determine the coverage of the filter.
Ideally, the filter would identify every document relevant to the
query and return it as part of the (unordered) answer set; which
suggests that it makes sense to measure the quality of the filter by
computing
\begin{equation}\label{eqn-recall}
	\metric{Recall}(\coll') 
		= \frac{|\coll' \cap \judg_q|}{|\judg_q|} 
\end{equation}
where $\coll' = F(\coll,\quer,k)$ is the set of $k$ documents the
filter extracted from the collection.
If a filter has a high recall according to Equation~\ref{eqn-recall},
then the set of documents passed to the ranking stage contains all or
most of those that the ranker could possibly preference, and so
retrieval effectiveness should not be degraded by the filter's
presence.
However, the converse is not true: a low recall score does not
necessarily give rise to a low score from the final effectiveness
metric, since the final metric might be strongly top-weighted, and
there might be many relevant documents for the filter to select
amongst.
Unfortunately, recall requires relevance judgments,
limiting to applicability to queries which have judgments available.

\myparagraph{Overlap}
If two filters are to be compared, the outputs
$\coll^a=F_a(\coll,q,k)$ and $\coll^b=F_b(\coll,q,k)$ can be tested
against each other, and an {\emph{overlap coefficient}} computed,
an approach that avoids the need for relevance judgments.
If it is assumed that the eventual evaluation metric {\metric{M}} is
insensitive to small changes in the sets, and $\coll_a$ and $\coll_b$
are close to being the same, then substituting $F_a()$ by $F_b()$ in
a {\typetwolevel} system should be plausible.
That is, the difference between the sets generated by two filters
might in some situations be a valid surrogate for the evaluation of
the eventual metric.
One possible overlap coefficient is given by the Jaccard similarity coefficient:
\begin{equation}
	\metric{Overlap}(\coll^a,\coll^b) =
	  \frac{|\coll^a \cap \coll^b|}{|\coll^a \cup \coll^b|} \, ,
	\label{eqn-overlap}
\end{equation}
which is zero when the two sets are disjoint, and is one if and only
they are identical.
A variation on the overlap computation replaces the denominator of
Equation~\ref{eqn-overlap} by
$\min\{|\coll^a|, |\coll^b|\}$.
Other coefficients are possible, including ones that are asymmetric
and compute the fraction of one set that is present in the other as a
{\emph{coverage ratio}}.

Overlap-based approaches are effective if the two sets are of
comparable size, and if the differences between them are small.
They also have the considerable benefit of not requiring that
relevance judgments be available.
But in the case of ranking systems it is also desirable to be able to
compare the outcomes obtained when the two sets might be of quite
different sizes.
For example, one filter might generate a subset of the collection
that is a tenth the size of another; what matters in this case is
what change arises in the downstream effectiveness score generated by
the final metric.
Given that most metrics are top-weighted, the fact that $\coll^a$ and
$\coll^b$ differ in size may be less important than
Equation~\ref{eqn-overlap} might suggest.

\myparagraph{Rank-Biased Overlap}
Top-weighted set overlap computations applicable to ordered sequences
have also been presented.
{\citet{wmz10acmtois}} describe a method they call
{\metric{Rank-Biased Overlap}}, or {\metric{RBO}}, which computes a
top-weighted overlap score between two sequences.
A {\emph{user model}} akin to that of the rank-biased precision
effectiveness metric is presumed, in which the reader scans from the
top of the two lists, determining the depth that they will view the
two sequences to according to a geometric probability distribution
governed by a parameter $p$.
The value of $p$ is the conditional probability of the user stepping
from depth $i$ to depth $i+1$ in the pair of sequences.
When $p$ is close to one, the user is ``patient'', and expects to
examine a relatively long prefix of the two sequences; when $p$ is
smaller, the user is modeled as only examining a relatively short
prefix.
The {\metric{RBO}} score between the two sequences is then computed
as a weighted average of overlap ratios:
\begin{equation}
\label{eqn-rbo}
\metric{RBO}(\coll^a, \coll^b)
= (1-p) \sum_{i=1}^\infty p^{i-1} \cdot
	\frac{|\coll^a_{1..i} \cap \coll^b_{1..i}|}%
		{i}\, ,
\end{equation}
where $\coll_{1..i}$ refers to the first $i$ elements in ordered
sequence $\coll$.

Like {\metric{Overlap}}, {\metric{RBO}} has the advantage of not
requiring that relevance judgments be available, and hence
experiments can be carried out automatically over large sets of
sample queries, allowing high levels of confidence in the measured
outcomes.
But note that RBO can only be used if the filter $F()$ generates an
ordered sequence of documents -- it cannot be applied if the filter
produces a set as its output.
Nor does it exactly match any particular eventual effectiveness
metric, although there is strong relationship between {\metric{RBP}}
and {\metric{RBO}}.
Being metric-agnostic can be thought of as being a strength of
several of these approaches, but also as a weakness.

Another top-heavy measure is Dice Top (\metric{DT}), introduced by
{\citet{msoh13acmtois}}.
{\citeauthor{msoh13acmtois}}\ primarily use {\metric{DT}} to
characterize features which improve effectiveness in a learning to
rank model, but it could also be used to compare the similarity
between any two result sets.

\myparagraph{End-to-End Effectiveness}
Another way of measuring the effectiveness of a filter is to
determine the extent to which the insertion of the filter alters the
metric score compared to the pure ranking scheme, assuming that the
filter extracts a subset $\coll'=F(\coll,q,k)$ of the collection for
subsequent processing.
That is, both the with-filter and without-filter systems are regarded
as being ``end to end'', and a statistical test is carried out using
pairs
\[
	\langle
	\metric{M}(R(\coll,q), \judg_q),
	\metric{M}(R(\coll',q), \judg_q)
	\mid
	q \in \quer
	\rangle
\]
A comparison using this methodology might, for example, seek to
demonstrate that the introduction of the filter allows rejection of
the hypothesis that ``the filter reduces the effectiveness scores
attained by $\epsilon$ or more'', for some small $\epsilon$ specified
as part of the experimental design.
Like the recall-based method (Equation~\ref{eqn-recall}), this
approach requires that relevance judgments be available.

\myparagraph{Maximized Effectiveness Difference (MED)}
Recent work by {\citet{tc14arxiv}} offers a further alternative, and
provides the basis for the evaluation in our experiments.
Given two documents sequences $\coll^a$ and $\coll^b$ and a chosen
metric $\metric{M}$, a set of judgments $\judg'$ is identified that
maximally separates the $\metric{M}$-scores for the two orderings.
The resulting value is the {\emph{maximum effectiveness difference}}
with respect to {\metric{M}}:
\begin{equation}
	\med{M}(\coll^a,\coll^b) =
		\max_{\scriptsize \judg \subseteq \coll^a \cup \coll^b}
		|\metric{M}(\coll^a,\judg)
		- \metric{M}(\coll^b,\judg)|
	\, .
\label{eqn-med}
\end{equation}
The methodology presented by {\citeauthor{tc14arxiv}} allows a set of
partial judgments $\judg'\subseteq\judg$ to be specified as an
additional constraint, plus a list of documents that may not be
members of $\judg$.
Computation of $\med{M}$ when {\metric{M}} is recall-independent is
straightforward {\citep{tc14arxiv}}, even when these additional
constraints are added.
To see the usefulness of this concept for evaluating two-level
retrieval systems, suppose that
$
	\coll^a = R(F(\coll,q,k),q)
$
and that
$
	\coll^b = R(\coll,q)
$
for some query $q$ and complete collection $\coll$.
That is, suppose that $\coll^b$ is the sequence generated by ranker
$R()$ without a filtering step, and $\coll^a$ is the sequence
generated by the ranking when $k$ documents are selected by the
filter $F()$.
Then $\med{M}(\coll^a,\coll^b)$, as defined by
Equation~\ref{eqn-med}, represents the maximum possible degradation
in measured effectiveness (according to metric {\metric{M}}) that
results from the insertion of $F()$ in to the processing pipeline.

For example, suppose that some query gives rise to the ranking
\[
	\coll^b = \langle 20, 45, 17, 11, 33, 29,
		18, 56, 72, 91, 54, 83, 22, \ldots \rangle
\]
and that the documents allowed by the filter give rise to 
the subsequence
\[
	\coll^a = \langle 20, 45, 17, 33, 29,
		56, 72, 91, 54, 22, \ldots \rangle \,.
\]
Suppose further that the metric being used to measure system
effectiveness is rank-biased precision $\metric{RBP}$ with the
parameter $p=0.8$.
The definition of $\metric{MED}$, and the definition of
$\metric{RBP}$, mean that $\med{RBP}$ arises when the documents that
are common to the two sequences are deemed to be non-relevant, making
$\RBP(\coll^a)=0$; and the documents that appear in $\coll^b$ only
are all deemed to be relevant, that is, with $r_i=1$.
In the case of the example rankings, that gives rise to the two
computations
\begin{eqnarray*}
\RBP0.8(\coll^b) &=& \RBP0.8(\langle 0,0,0,1,0,0,1,0,0,0,1,\ldots\rangle)\\
	&=& 0.2(0.8^{3} + 0.8^{6} + 0.8^{10})\\
	&=& 0.176 \\
\RBP0.8(\coll^a) &=& \RBP0.8(\langle 0,0,0,0,0,0,0,0,\ldots\rangle) \\
	&=& 0.0 \, .
\end{eqnarray*}
So, $\med{RBP0.8}(\coll^a,\coll^b) = 0.176$, and is an upper
bound on the effectiveness loss when the filtered ranking $\coll^a$
is used to approximate the full ranking $\coll^b$ when the runs are
scored using $\RBP$ with $p=0.8$.
Note that no judgments are required in this computation; it is
derived solely from the full and filtered sequences of document
numbers.
Moreover, it can be based on any recall-independent effectiveness
metric, meaning that in many situations it will be possible to employ
the metric that is used to measure end-to-end system effectiveness.

\citet{tc14arxiv} indicate that the $\med{}$ family of metrics was directly
inspired by an attempt to generalize $\metric{RBO}$.
In the same way that there is a close association between $\metric{RBP}$
and $\metric{RBO}$,
$\med{}$ provides a general procedure for deriving a rank similarity
measure from a (recall independent) effectiveness measure.
\citet{tc14arxiv} show that $\med{RBP}$ is highly correlated with
$\metric{RBO}$, supporting our choice of the $\med{}$ family
for measuring filter-stage effectiveness.

\section{Filters}
\label{sec-filters}

We next describe the various filters that might be employed in a
staged retrieval system, and then in Section~\ref{sec-validation}
present validation experiments that confirm the relationship between
$\med{M}$ and the effectiveness loss (relative to metric
$\metric{M}$) that arises when filtering is employed.
Section~\ref{sec-tradeoffs} steps away from the use of relevance
judgments, and expands the experimentation to a large query sequence.
A key goal is to demonstrate that the $\med{M}$ approach yields
usable information about filtering-stage effectiveness, but we are
also able to compare and contrast the relative efficiency of the
various filtering mechanisms.

\myparagraph{Pure Boolean Conjunction}
Conjunctive Boolean queries have been extensively studied by the IR
research community {\citep{cm10-tois,kt14sigir}}.
Using hybrid postings lists that combine bitvectors and compressed
postings can significantly improve the efficiency in terms of both
time and space.
Efficiency can be further improved using document reordering
{\citep{kt14sigir}}.
However, the fastest of these approaches cannot use the same inverted
index for ranked querying and Boolean filtering.

\newcommand{\bsuccessor}{\mbox{\textbf{\textsc{Successor}}}}
\newcommand{\bfsearch}{\mbox{\textbf{\textsc{F-Search}}}}
\newcommand{\bid}{\mbox{\textbf{\textsc{Id}}}}
\newcommand{\bsort}{\mbox{\textbf{\textsc{Sort}}}}
\newcommand{\bappend}{\mbox{\textbf{\textsc{Append}}}}
\newcommand{\bmini}{\mbox{\textbf{\textsc{MinId}}}}

\begin{algorithm}[t]
\raggedright
{\textsc{input}}: A list of $q$ ordered sets $S_1 \ldots S_{|q|}$. 

{\textsc{output}}: An ordered set of documents $R$.

\begin{algorithmic}[1]
\State set $R \leftarrow \varnothing$
\State initialize each $S_i$ with its least element as its candidate
\State \bsort$(S_1\ldots S_{|q|})$ 
    based on candidate
\State set $x \leftarrow \mbox{the candidate from $S_{|q|}$}$
\While{$x$ is defined} 
    \If{the candidate from $S_1$ is equal to $x$}
	    \State \bappend$(R, x)$
	    	\label{stp-append}
    \EndIf
    \State set $x \leftarrow {\bsuccessor}(S_{|q|},x)$
    \For{$i=1$ to $|q|-1$}
	    \State ${\bfsearch}(S_i, x)$
    \EndFor
    \State \bsort$(S_1\ldots S_{|q|})$ 
        based on candidate 
    \State set $x \leftarrow \mbox{the candidate from $S_{|q|}$}$
\EndWhile
\Return{$R$}
\end{algorithmic}
\mycaption{Boolean Intersection Algorithm ({\small\sf{Boolean}})}
\label{alg-bool}
\end{algorithm}

The experiments reported below make use of two different Conjunctive
Boolean Algorithms.
The first variant is the {\tt{\#band}} operator in
{\indri}\footnote{\small\url{http://www.lemurproject.org/indri/}} which
performs a full Boolean pass over the index.
We also implement a second variant based on the Adaptive Intersection
Algorithm {\citep{dlom00-soda,dlom01-alenex}} which is easily
amenable to a block-compressed inverted index, and can be processed
in a document-at-a-time manner similar to WAND.
The algorithm used is shown in Algorithm~{\ref{alg-bool}}.
The processing is similar in spirit to WAND.
First the smallest document identifier is selected as the target identifier $x$, and
all lists are sorted in increasing order by the current document id
cursor.
If the first cursor and last cursor are equal, the identifier is appended to
the results list.
Next, the target identifier $x$ is set as the successor from the last
postings list, and the cursors in every list are forwarded to $x$, or
its successor.
When the end of any postings list is reached, the algorithm
terminates.

In contrast to the term-at-a-time based intersection algorithm SvS
used by {\citet{al13sigir}}, this intersection algorithm is directly 
amenable to document-at-a-time scoring algorithms such as WAND.
If a full conjunctive Boolean result set is desirable, using a SvS approach
is the most efficient processing scheme in practice~\citep{cm10-tois}.
However, SvS does not easily support early termination after $k$ items
have been identified, but a WAND-like traversal of the lists does.

\myparagraph{Boolean Conjunction with Static Scores}
The effectiveness of Boolean conjunction as a filtering stage can be
enhanced by the inclusion of a static scoring process.
This is commonly achieved by reordering the document collection using a
static score before constructing the index.
Common static scoring regimes include spam score {\citep{csc11-ir}},
PageRank {\citep{page1999}}, or document length.
If $k$ documents are to be retrieved, the first $k$ that are
identified at step~\ref{stp-append} of Algorithm~{\ref{alg-bool}} are
the ones returned -- they will have the highest static score amongst
all documents that contain all of the query terms.

It is also possible to integrate on-the-fly static scoring into
Algorithm~{\ref{alg-bool}} by adding a min-heap structure as part of
the {\bappend} operation at step~\ref{stp-append}.
The advantage of this arrangement is that the same index can be used
to evaluate a range of different static scores; the drawback, of
course, is that processing cannot be terminated after the first $k$
matching items have been determined.
Unless otherwise noted, we assume here that the collection is
pre-ordered according to static score, and that the first $k$ that
contain all of the query terms according to a single static scoring
method are the ones passed through to the ranking stage.


%

%




\myparagraph{WAND and MaxScore}
A third filtering option to use is an efficient bag-of-words scoring
algorithm such as WAND (Weak {\sc{and}}, or Weighted {\sc{and}})
{\citep{zchsz03cikm}} or MaxScore {\citep{tf95ipm,stc05-sigir}}.
Both algorithms can be used with a variety of ranking functions,
including the Okapi BM25 computation and methods based on language
models.
{\citet{mso13irj}} document the advantages and disadvantages of using
a bag-of-words filtering step for second stage learning to rank
algorithms.
{\citet{al13sigir}} and {\citet{wlm11sigir}} also investigate various
trade-offs with bag-of-words filtering in multi-stage retrieval
architectures.

\myparagraph{Aggressive WAND}
The WAND algorithm is defined as follows.
Given a list of Boolean indicator variables $\mn{x_1, x_2,\ldots,
x_q}$ with $x_i\in\{0,1\}$ indicating whether or not the $i$\,th term
in the query appears in the current document, a list of positive
weights $\mn{U_1},\mn{U_2},\ldots,\mn{U_q}$ with $U_i$ indicating the
largest score contribution that the $i$\,th term in the query gives
rise to in any document in the collection, and a threshold $\mn{t}$,
\[
\mbox{\sc{wand}}\mn{(x_1,U_1,\ldots, x_q, U_q, \mn{t})}
	\equiv
	\left(
	\sum_{1 \le \mn{i} \le \mn{q}} \mn{x_i U_i} \ge \mn{t}
	\right)
	\, .
\]
Therefore, $\mbox{\sc{wand}}$ is a Boolean indicator that is $0$ when
there is no possibility that the score for the current document can
exceed $t$, and $1$ if and only if it can.
Let $s_{\mbox{\scriptsize{min}}}$ be the value of the $k$\,th largest
document score that has been generated to date in the computation.
In the {\emph{aggressive WAND}} strategy the indicator
\[
\mbox{\sc{wand}}\mn{(x_1,U_1,\ldots, x_q, U_q, \theta \cdot
	s_{\mbox{\scriptsize{min}}})}
\]
is used to provide guidance as to whether the current document should
have its score computed.

If $\theta = 1$ then a standard evaluation takes place, and every
document that {\emph{might}} be able to enter the heap will be
scored, with the overall top-$\mn{k}$ items guaranteed to be
returned.
When $\theta > 1$, the barrier is raised, and fewer documents are
scored, on the presumption that any documents that have scores that
are close to the heap's lower limit at the time they are first
encountered are likely to be evicted through the course of the
remaining processing.
The risk is that there is one or more documents with WAND estimates a
little above $s_{\mbox{\scriptsize{min}}}$ at the time they are
processed, plus have actual similarity scores also greater than
$s_{\mbox{\scriptsize{min}}}$; and that the computation as a whole
has stabilized, so that $s_{\mbox{\scriptsize{min}}}$ does not rise
very much subsequently, and this document would not have been
displaced.
The larger $\theta$, the greater the risk that such errors might
occur.
In the limit, if $\theta = \infty$, only the first $\mn{k}$ documents
evaluated get added to the heap, and no other documents will be
scored.
The experiments reported in the next section consider several
$\theta$ values greater than one, exploring the balance between
filtering effectiveness and processing efficiency.
In their paper describing WAND, {\citet{zchsz03cikm}} also note that
$\theta$ might be used as a parameter to accelerate searching while
risking the score-safe nature of the computation.
One of the contributions of this paper is a detailed exploration of
the usefulness of the aggressive WAND approach.

\myparagraph{Scored Boolean WAND}
Another variation on WAND is to only Okapi-score documents which are
a full conjunctive match, and contain all of the query terms.
This is implemented as a modification to the Adaptive Intersection
algorithm shown in Algorithm~{\ref{alg-bool}}, with the {\bappend}
operation at step~\ref{stp-append} replaced by first an evaluation of
the Okapi score of that document, then a test against the entry
threshold for a min-heap containing the $k$ largest scores accumulated
for documents processed to this moment, and then finally, if the new
score is greater, suitable heap operations to update the state of the
heap.
This variation on WAND has a similar effectiveness profile to the
{\tt{\#band}} operator in {\indri}, which is not optimized for
performance.
In general, $\theta$ has less of an effect on scored Boolean WAND
than it does in the standard Okapi computation, since the fact that
all of the query terms must always appear means that the sum of the
upper bound scores in pivot evaluation is always relatively high.

\myparagraph{Other Approaches}
%
%
{\citet{wf14cikm}} present another compromise between Scored Boolean
and Static Boolean.
The key idea is to use a decision tree with per term IDF values to
prioritize the matches.
The efficiency and effectiveness horizon of this approach lies
between the two methods explored in this work.
{\citet{al13sigir}} also investigate the impact of disjunctive and
conjunctive WAND when used as a filtering step for a learning-to-rank
second stage derived from Lambda Mart.
{\citeauthor{al13sigir}} conclude that conjunctive WAND was the best
compromise as the final effectiveness results were statistically
indistinguishable from the disjunctive WAND filter.
But they do not consider aggressive WAND variations, or the impact of
alternative second stage runs.
Finally, {\citet{wlm11sigir}} present a cascaded learning-to-rank
approach to iteratively shrink the candidates evaluated.
The approach is efficient and effective in practice, but relatively
difficult to separate into distinct stages for independent
evaluation.


\section{Validation Experiments}
\label{sec-validation}

\myparagraph{Experimental Setup}
In order to validate the use of {\method{MED}} to establish a
correlation between filtering effectiveness and end-to-end
effectiveness, a detailed evaluation using judged topics has been
undertaken.
All of the experiments described in this section and the next section
use Part B of the ClueWeb 2009 collection (CW09B); in this section
(only), the 2010 and 2012 Ad-Hoc queries from the TREC Web Track are
also used ($50$ queries per year), together with the relevance
judgments associated with them.
We refer to these collections of topics as AH2010 and AH2012
respectively.

To ensure our effectiveness results are consistent with commonly used
search engines, we use version $5.6$ of {\indri}.
For {\indri} Boolean runs, we use the {\tt{\#band}} operator; for
Language model runs we use the default parameters; and for Okapi BM25
runs we use $k_1=0.9,b=0.4,k_3=0$, since these values consistently
give better results than the default parameters {\citep{tjc12-osir}}.
All runs employ stemming using the Krovetz stemmer, and the default
stopwords list is used when stopping is enabled.
Runs without stopping were also carried out.

\begin{table}[t]
\centering
\newcommand{\tabent}[1]{\begin{minipage}[t]{103mm}\raggedright{#1}~\\[-0.5ex]~\end{minipage}}
\begin{tabular}{lc}
\toprule

Stage One Run Methods & Runs
\\

\midrule

\tabent{Bag-of-words, language model, Dirichlet smoothing (\indri),
	$\times$~stopped or not, $\times$~spam-pruned or not}
	& 4
\\
\tabent{Bag-of-words, BM25, $k_1=0.9,b=0.4$ (\indri),
	$\times$~stopped or not, $\times$~spam-pruned or not}
	& 4
\\
\tabent{Bag-of-words, aggressive WAND, BM25, $k_1=0.9,b=0.4$,
	stopped, not spam-pruned,
	$\theta \in \{1.0, 1.02, 1.05, 1.1, 1.2, 1.5, 2.0\}$}
	& 7
\\
\tabent{Scored Boolean, WAND, BM25, $k_1=0.9,b=0.4$,
	only used in Section~\ref{sec-tradeoffs}}
	& 0
\\
\tabent{Scored Boolean, {\indri} {\tt{\#band}},
	$\times$~stopped or not, $\times$~spam-pruned or not}
	& 4
\\
\tabent{Boolean, static pagerank as raw scores,
	$\times$~stopped or not, $\times$~spam-pruned or not}
	& 4
\\
\tabent{Boolean, static pagerank as bucketed log probability prior,
	$\times$~stopped or not, $\times$~spam-pruned or not}
	& 4
\\
\tabent{Boolean, static fusion spam score,
	$\times$~stopped or not, $\times$~spam-pruned or not}
	& 4
\\[-1.0ex]
\bottomrule

\end{tabular}

\mycaption{Summary of filter stage mechanisms explored.
A total of four different {\indri} indexes were constructed from
CW09B for each filter type: an unstopped index, a stopped index, an
unstopped index containing documents with spam scores greater than
$70$ only, and a stopped index containing documents with spam scores
greater than $70$ only.
In the three static Boolean approaches, document length was used as a
secondary key.
A total of $6\times 4 + 7=31$ filtering mechanisms were tested in the
experiments reported in this section.
\label{tbl-stage1}}
\end{table}

\myparagraph{Results}
%
Table~{\ref{tbl-stage1}} summarizes the filter-stage combinations used
for the validation and efficiency experiments.
For all of the filter stage runs, we investigated four different types of
indexes: Unstopped and Unpruned; Stopped and Unpruned; Unstopped and
Pruned; and, fourthly, Stopped and Pruned.
The pruned indexes were restricted to only include documents with
spam scores greater than $70$, using the {\emph{Fusion}} scores
described by {\citet{csc11-ir}}.
Each of the listed filter-stage options was evaluated to depths $k$
of $20$, $50$, $100$, $200$, $500$, $1{,}000$, $2{,}000$, $5{,}000$,
and $10{,}000$.

Two different orderings by PageRank were also used.
The first version used unnormalized raw PageRank scoring; the second
version used the binned log probability values.
Both versions used are freely available on the ClueWeb09
Wiki\footnote{\small\url{http://boston.lti.cs.cmu.edu/clueweb09/wiki/tiki-index.php?page=PageRank}}.

For gold standard final ranking stage runs, we employed the top runs
submitted to TREC for the corresponding query sets.
These are listed in Table~{\ref{tbl-stage2}}.
For AH2010 and AH2012, the selected runs achieved the best
performance over Part~B of the ClueWeb 2009 collection under the
primary measure used for reporting track results ($\metric{ERR@20}$).



\begin{figure}[p]
\centering
\showgraph{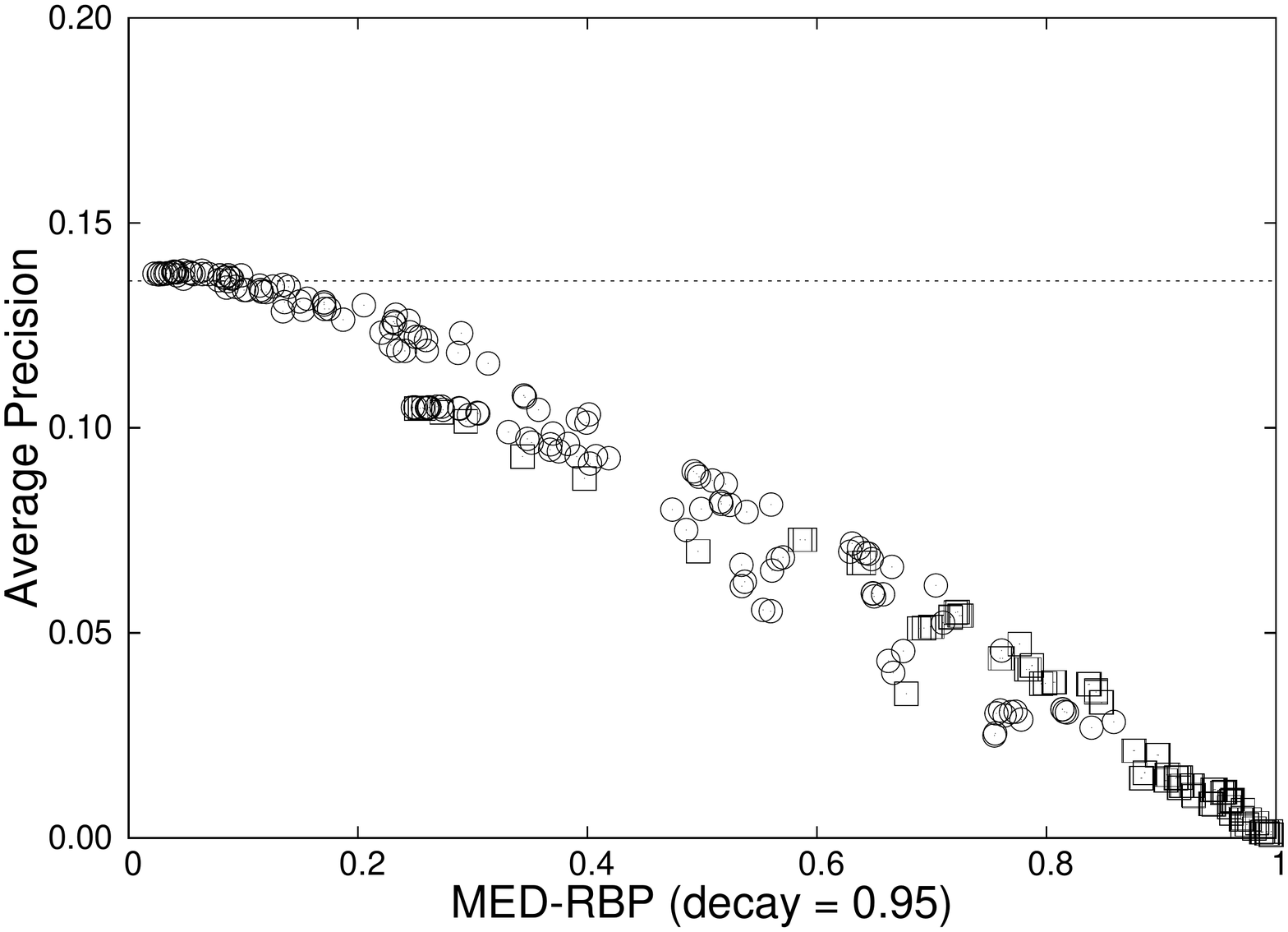}
\showgraph{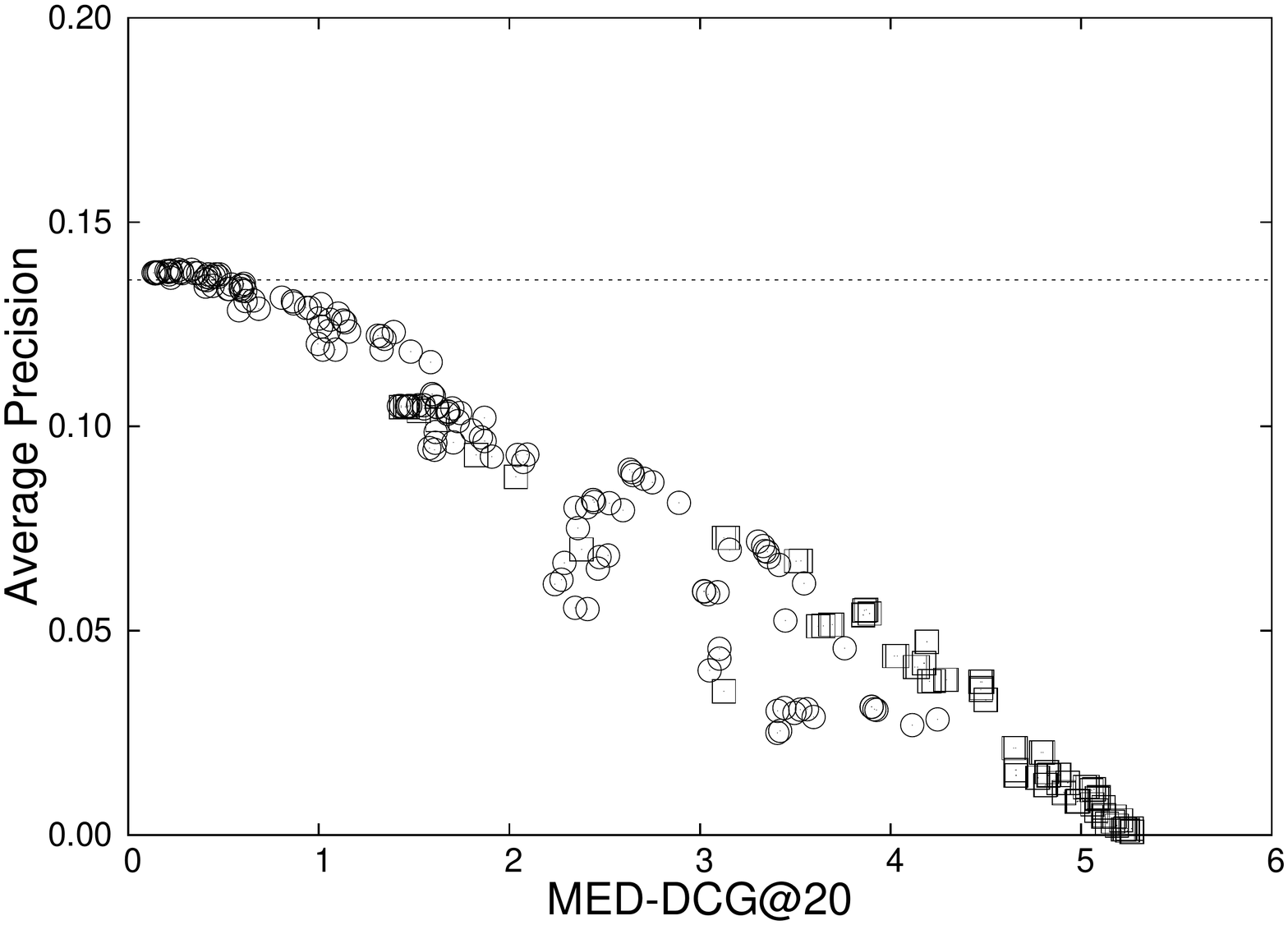}
\caption{
Correlations between $\med{RBP0.95}$ and $\med{DCG20}$ and
measured $\AP$ using the AH2010 collection, $50$ queries, and
{\tt{IvoryL2Rb}} as the final ranking stage.
Each of the $279$~points represents one of $31$ distinct filter
stages employed at one of nine filter-stage depths, ranging from $k =
20$ to $k = 10{,}000$.
The dashed line indicates the performance of the final stage with no
early-stage filter.
Circles indicate early-stage filters that use query-dependent ranked
retrieval; squares indicate early-stage filters that use Boolean
retrieval with static ranks.
\label{fig-validation1}}
\end{figure}

\begin{table}[t]
\centering
\begin{tabular}{llc}
\toprule

Collection & TREC Run Identifier & Depth\\

\midrule


MQ2009 & {\tt uogTRMQdph40} & $\D1{,}000$ \\
AH2010 & {\tt IvoryL2Rb}  & $10{,}000$ \\
AH2012 & {\tt DFalah121A} & Variable depth ($d = 11$ -- $2{,}009$) \\

\bottomrule

\end{tabular}

\mycaption{
Second stage runs used as reference points.
\label{tbl-stage2}}
\end{table}

\begin{figure}[p]
\centering
\showgraph{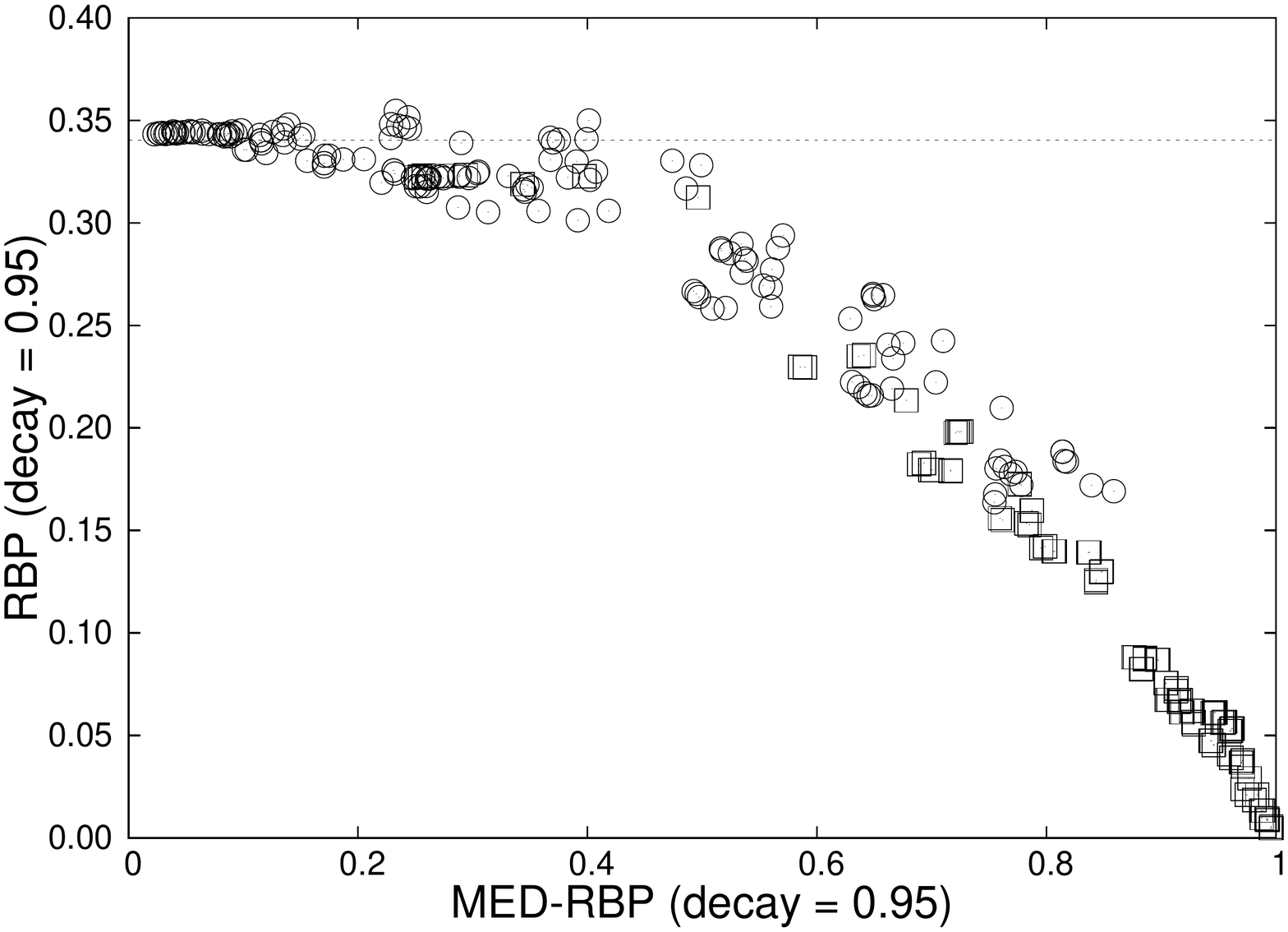}
\showgraph{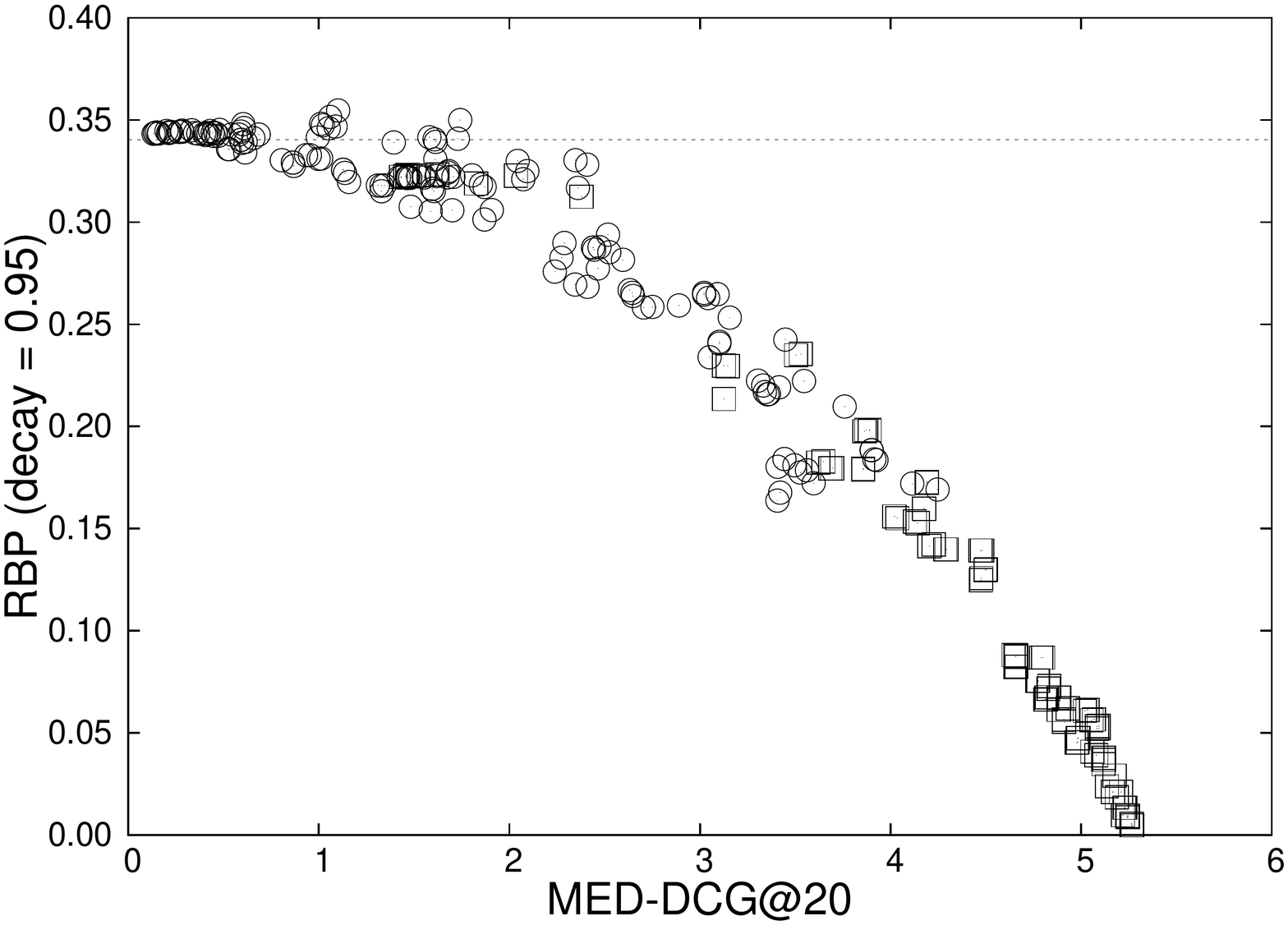}
\caption{
Correlations between $\med{RBP0.95}$ and $\med{DCG20}$ and
measured {\metric{RBP0.95}} using the AH2010 collection, $50$ queries, and {\tt
IvoryL2Rb} as the final ranked stage.
Each of the $279$~points represents one of $31$ distinct filter
stages employed at one of nine filter-stage depths, ranging from $k =
20$ to $k = 10{,}000$.
The dashed line indicates the performance of the final stage with no
early-stage filter.
Circles indicate early-stage filters that use query-dependent ranked
retrieval; squares indicate early-stage filters that use Boolean
retrieval with static ranks.
\label{fig-validation1a}}
\end{figure}

\begin{figure}[t]
\centering
\showgraph{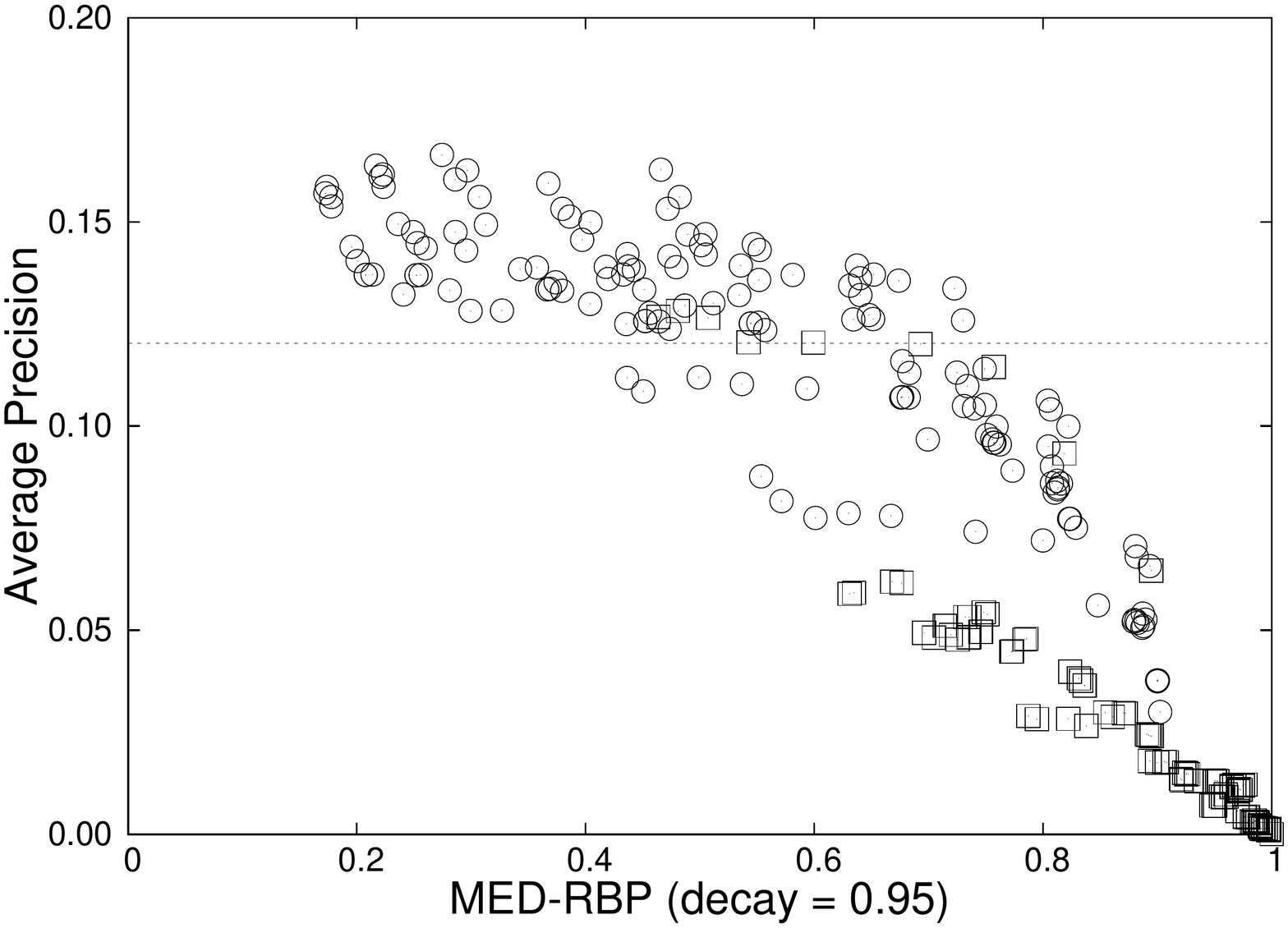}
\caption{
Correlation between $\med{RBP0.95}$ and measured $\AP$ using the
AH2012 collection, $50$ queries, and {\tt{DFalah121A}} as the final
ranked stage.
Each of the $279$~points represents one of $31$ distinct filter
stages employed at one of nine filter-stage depths, ranging from $k =
20$ to $k = 10{,}000$.
The dashed line indicates the performance of the final stage with no
early-stage filter.
Circles indicate early-stage filters that use query-dependent ranked
retrieval; squares indicate early-stage filters that use Boolean
retrieval with static ranks.
\label{fig-validation2}}
\end{figure}

The top plot of Figure~\ref{fig-validation1} shows the correlation
between measured $\med{RBP0.95}$ and measured $\AP$.
The bottom plot of Figure~\ref{fig-validation1} shows the correlation
between measured $\med{DCG20}$ and measured $\AP$.
These plots cover a suite of $31$ different filter stages, and nine
different depths $k$ for each first stage.
For this experiment, the measures are computed over the 50~queries of
the AH2010 collection, with {\tt{IvoryL2Rb}} forming the gold
standard final-stage ranked run, assumed
to reorder the documents provided by the filter phase
according to the similarity score computed by the original TREC run.
The performance of the final stage with no early-stage filtering is
shown by the line at $0.1358$.
The diagonal lines demonstrate the clear inverse relationship between
end-to-end $\AP$ and, respectively, $\med{RBP0.95}$ and $\med{DCG20}$
In this figure the smooth transition from good overall performance to
bad overall performance suggests that the good filter-phase options
are well-matched to the final stage computation that is being used.

Early-stage filters with $\med{RBP0.95}$ values under $0.15$
(or $\med{DCG20}$ values under $0.75$) have
little impact on end-to-end effectiveness as measured by $\AP$.
These early-stage filters are based on bag-of-words ranked retrieval
methods (for example, BM25) and large retrieval depths (for example,
$k = 10{,}000$).
Their low $\med{}$ values suggest that they are providing the
final stage with an appropriate set of documents for re-ranking,
even though the reranking stage incorporates ranking methods and
features not found in these filter stages, including a learned ranker
and term proximity~\citep{ivory10}.
With smaller values for $k$, the $\med{}$ values of these
filters increase as their end-to-end performance decreases.
Filter stages with $\med{RBP0.95}$ values over $0.90$
(or $\med{DCG20}$ values over $4.5$) use Boolean
retrieval and smaller values of $k$.

While filter stages with low $\med{}$ values provide
essentially the same end-to-end effectiveness,
their low $\med{}$ values serve to differentiate them,
with lower values suggesting a better fit with the final stage ranker.
The $\AP$ values are not able to provide this differentiation,
possibly due to the presence of unjudged documents,
which are assumed to be non-relevant.
By treating the output of the final stage as a gold standard and directly
measuring the impact of filtering in an early stage,
we avoid the complexities introduced by these unjudged documents.

Figure~\ref{fig-validation1a} shows equivalent plots for measured
$\RBP$.
The flattening towards the left suggests that $\metric{RBP0.95}$,
which is a shallower effectiveness metric than {\AP}, is less
sensitive than $\AP$ to distinctions between filtering stages.
While we do not show plots, similar correlations also arise when
other effectiveness measures are used with $\med{}$, including
{$\metric{NDCG@20}$}, {$\metric{P@10}$}, and {$\metric{ERR}$}.
Effectiveness measures focused on precision at low ranks,
particularly ${\metric{ERR}}$, are less sensitive to the choice of
early-stage rankers, with wide ranges of $\med{}$ values
corresponding to similar measured end-to-end effectiveness.
This relative tolerance is as expected -- shallow end-to-end metrics
can be satisfied by rankings with more divergence than can deep
end-to-end metrics, since the second ranking stage can still find the
documents it needs, even if further down the first-phase's output,
and get them to the top of a ranking for the shallow metric to
benefit from.
That is, if a deep-weighted $\med{}$ evaluation is carried out as a
predictor of a shallow end-to-end metric, higher $\med{}$ scores can
be regarded as being acceptable.




Figure~\ref{fig-validation2} shows a more complex situation.
For this experiment, the measures are computed over the 50~queries of the
AH2012 collection.
The run {\tt DFalah121A} forms the final ranking stage,
chosen for its excellent performance on the 50~queries of AH2012 when
measured by $\metric{ERR}@20$~\citep{ottawa12}.
This run took unusual approaches to indexing and query processing,
making substantial use of external resources.

There are a number of points to be noted.
First, none of the early-stage filters generates the documents that
the final stage is wishing to see in the top-ranked positions, and
the result is that $\med{RBP0.95}$ is never less than about $0.2$.
Second, this discrepancy does not greatly harm the measured $\AP$
score for the combined run.
Indeed, the best early-stage filters actually increase $\AP$
substantially above the score of the ranking stage without filtering,
as shown by the line at $0.1203$.
These filtering stages are essentially forcing the ranking stage to
improve under the recall-based $\AP$ effectiveness measure, by
restricting the set of the documents it can rank.
Third, the correlation between $\med{RBP0.95}$ and measured $\AP$ is
weaker than in Figure~\ref{fig-validation1}, again suggesting that
these early stage filters may be a poor fit with this final stage.

While the improvements to $\AP$ and the weaker correlation would not be
visible without relevance judgments,
the poor fit would be noticeable from the lack of low $\med{RBP0.95}$ values.
This example emphasizes an important limitation of our approach.
We assume that the final ranking stage provides an acceptable gold standard for
comparing early-stage filters.
Poor end-to-end performance cannot be detected.

\section{Judgment-Free Measurement of Effectiveness Tradeoffs}
\label{sec-tradeoffs}

We now explore the relationship between filter-stage retrieval
effectiveness and efficiency using a large query log, and $\med{RBP}$
as a measure for filtering effectiveness.


\myparagraph{Experimental Resources}
For the efficiency experiments, $40{,}000$ queries from the 2009
Million Query Track were used with an stopped and unpruned CW09B
index.
We refer to this combination as MQ2009.
The {\tt{uogTRMQdph40}} system is used as the gold standard, as it
represents the top-scoring system (when measured over the small
subset of the queries that were evaluated) that returned runs for all
of the large set of $40{,}000$ MQ2009 queries.
Algorithms were implemented in C++ and compiled with {\method{gcc
4.8.1}} using {\method{--O3}} optimizations; and experiments were run
on a 24 core Intel Xeon E5-2630 running at 2.3~GHz using {\gb{256}}
of RAM.
All efficiency runs are reported as the median of the per query
execution times of a single execution of a complete stream containing
one instance of each query in MQ2009, executed entirely in-memory.
Postings lists are stored compressed using the FastPFOR library
{\citep{lb-spe15}}, with skipping enabled.
As discussed in the previous section, each of the listed filter-stage
options was evaluated to depths $k$ of $20$, $50$, $100$, $200$,
$500$, $1{,}000$, $2{,}000$, $5{,}000$, and $10{,}000$.

%

\begin{figure}[t]
\centering
\showgraph{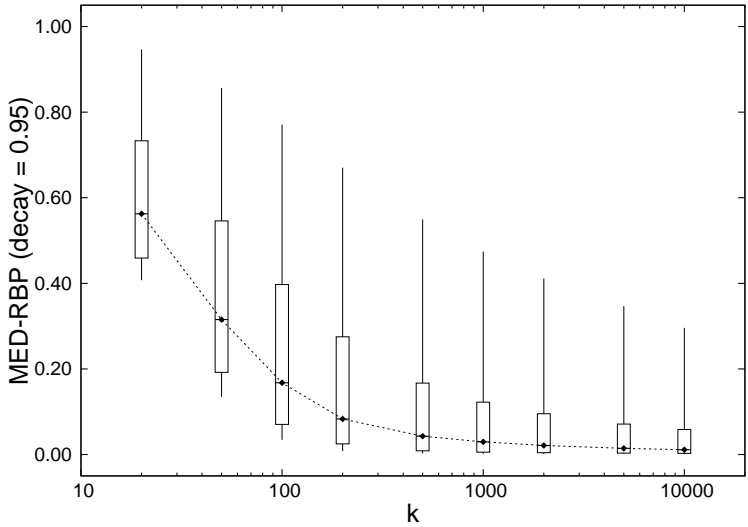}
\caption{Okapi ranking as early stage filter to retrieve $k$
documents, then using {\tt{uogTRMQdph40}} as the final ranking stage
to order them using the stopped unpruned CW09B collection.
Boxplots show $\med{RBP0.95}$ values over the $40{,}000$ MQ2009
queries, with boxes extending from the first to the third quartile
and with whiskers extending to the $10$\,th and $90$\,th percentiles.
Median values are connected by lines.
\label{fig-boxplot1}}
\end{figure}

\begin{figure}[t]
\centering
\showgraph{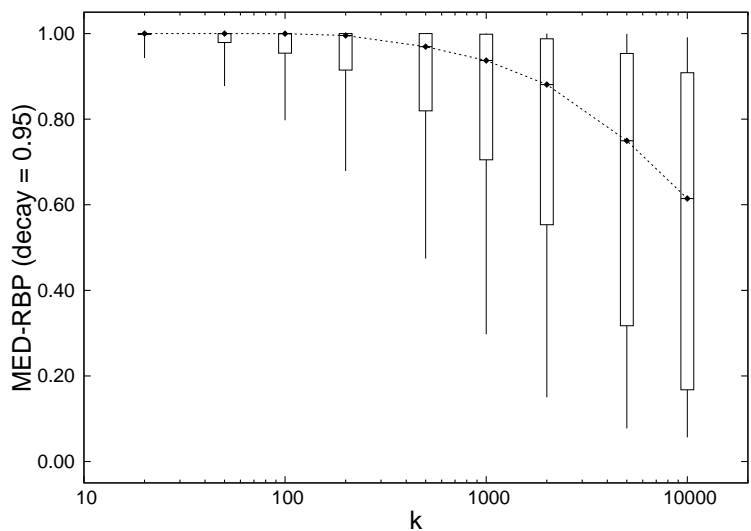}
\caption{Boolean conjunction with the collection pre-ordered by
PageRank as the filter stage to retrieve $k$ documents, then using
{\tt{uogTRMQdph40}} as the final ranked stage to order them.
Boxplots show $\med{RBP0.95}$ values over the $40{,}000$ MQ2009
queries, with boxes extending from the first to the third quartile
and with whiskers extending to the $10$\,th and $90$\,th percentiles.
Median values are connected by lines.
\label{fig-boxplot2}}
\end{figure}

\myparagraph{Filter Stage Effectiveness}
Figures~\ref{fig-boxplot1} and~\ref{fig-boxplot2} demonstrate
typical outcomes.
In each of the two plots, $\med{RBP0.95}$ is used as the
effectiveness assessment, and the depth $k$ of the filter-phase
output is plotted on the horizontal axis.
The reference run in both cases came from the University of Glasgow
(run {\tt{uogTRMQdph40}}).

In Figure~\ref{fig-boxplot1}, which makes use of an Okapi similarity
computation for the filtering phase, there is a clear inverse
relationship between $k$ and the upper bound on quality.
Passing as few as $1{,}000$ documents to the final ranking phase is
sufficient to obtain a $\med{RBP0.95}$ score of $0.2$, which is
suggestive of good quality final outcomes
(Figure~\ref{fig-validation1}).
Nonetheless, Figure~\ref{fig-boxplot1} shows a wide variance of
$\med{RBP0.95}$ scores across the set of queries.
While the median (marked by the overlaid line) shows a clear trend,
there are upper outlier queries for which even deep filtering to
depth $10{,}000$ is still inadequate.
For example, ``{\tt{buying first home}}'' (query \#50066) has a
$\med{RBP0.95}$ value of one.
Okapi scores for the top $10{,}000$ documents fall into the narrow
range $[3.75338,3.86595]$, suggesting that the collection includes
many documents containing these terms in similar proportions.
None of these $10{,}000$ documents are ranked above $1{,}000$ by the
final ranking stage.

On the other hand, Figure~\ref{fig-boxplot2} shows that filtering
based on Boolean conjunction and a static PageRank score is unlikely
to give good effectiveness in a multi-stage retrieval system,
regardless of the number of documents identified by in the filtering 
stage.
That is, the presence of all of the query terms alone is of limited
usefulness towards identifying potentially relevant documents,
even when coupled with PageRank as an ordering criteria.
Again, however, there are outlier queries that are at odds with this
overall trend.
The query ``{\tt{basscat boats}}'' (\#36318) provides one example.
Only $102$ documents contain both terms, so that even at depth $100$
the $\med{RBP0.95}$ score falls close to zero.


Across the set of filter-stage methods explored, there was a
consistent separation -- they either gave plots that corresponded to
Figure~\ref{fig-boxplot1}, or they gave plots similar to
Figure~\ref{fig-boxplot2}.
The differentiating criteria was very simple.
Methods based on Boolean conjunction, even when combined with a
static pre-ordering criteria such as PageRank, gave uniformly poor
results, whereas methods that involved an element of ranking performed
as shown in Figure~\ref{fig-boxplot1}, with a gradated trade-off
between $k$ and measured effectiveness.
Within the latter group there were small difference in
$\med{RBP0.95}$.
For example, aggressive WAND processing with $\theta=2.0$ was less
effective (or rather, gave higher $\med{RBP0.95}$ values) than when
smaller values of $\theta$ were used.
This outcome is consistent with our validation experiments in the
previous section
(Figures~\ref{fig-validation1},~{\ref{fig-validation1a}},
and~\ref{fig-validation2}) where these Boolean filters (marked as
squares in the plots) exhibited poor performance.

\begin{figure}[t]
\centering
\showgraph{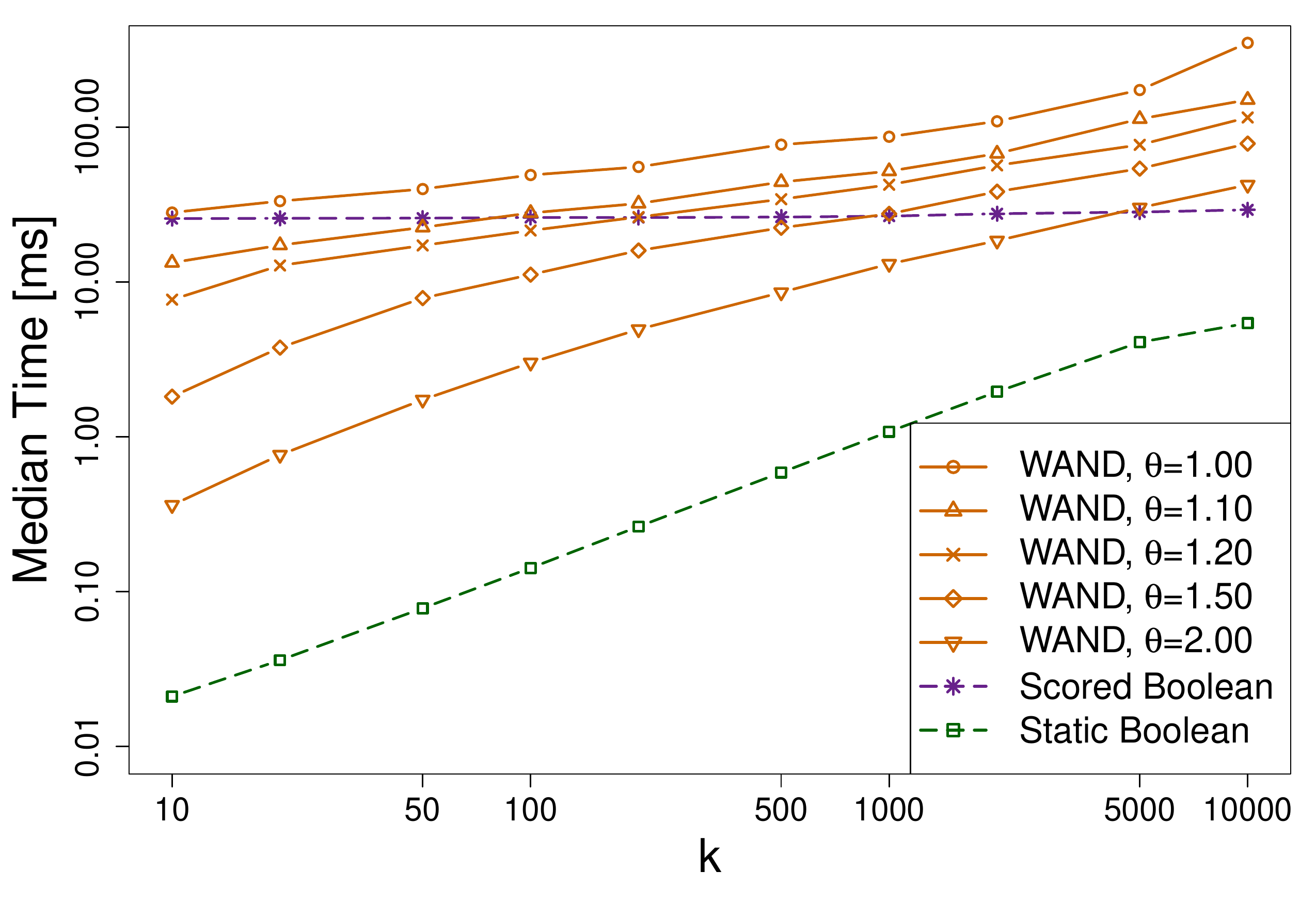}
\caption{Median query execution time in milliseconds for fully
in-memory execution, plotted as a function of $k$, the number of
documents required, using $40{,}000$ queries and the stopped unpruned
CW09B collection.
\label{fig-firststagespeed}}
\end{figure}

\myparagraph{Filtering Stage Evaluation Cost}
Figure~\ref{fig-firststagespeed} shows the flip side of those
effectiveness results, plotting filter-stage evaluation time as a
function of $k$.
Strict conjunctive Boolean evaluation is extremely fast, partly
because (using Algorithm~\ref{alg-bool}) the processing required per
answer document is modest, and partly because filter-stage evaluation
can be abandoned just as soon as $k$ matching documents have been
identified (an option not shown in Algorithm~\ref{alg-bool}, but
trivial to implement).
The execution time trends shown for the WAND variants are reflected
in operation counts for document scoring, heap insertions, and so on,
confirming the basis of the time savings.
The fastest of the aggressive WAND strategies, using $\theta=2.0$, is
around an order of magnitude slower than pure Boolean filtering -- at
least one of the postings lists for the terms must be fully scanned
before any documents can be returned at all, and the per document
processing cost is also higher.
That aggressive WAND variant is in turn is another one to two orders
of magnitude faster than the set-safe $\theta=1.0$ WAND version.

The scored Boolean WAND method provides a different computational
profile to the aggressive WAND implementations.
Regardless of $k$, it performs roughly the same amount of work in
terms of scoring; and the reduced number of heap operations when $k$
is small is not enough to have a major influence on execution cost.

\begin{figure}[t]
\centering
\showgraph{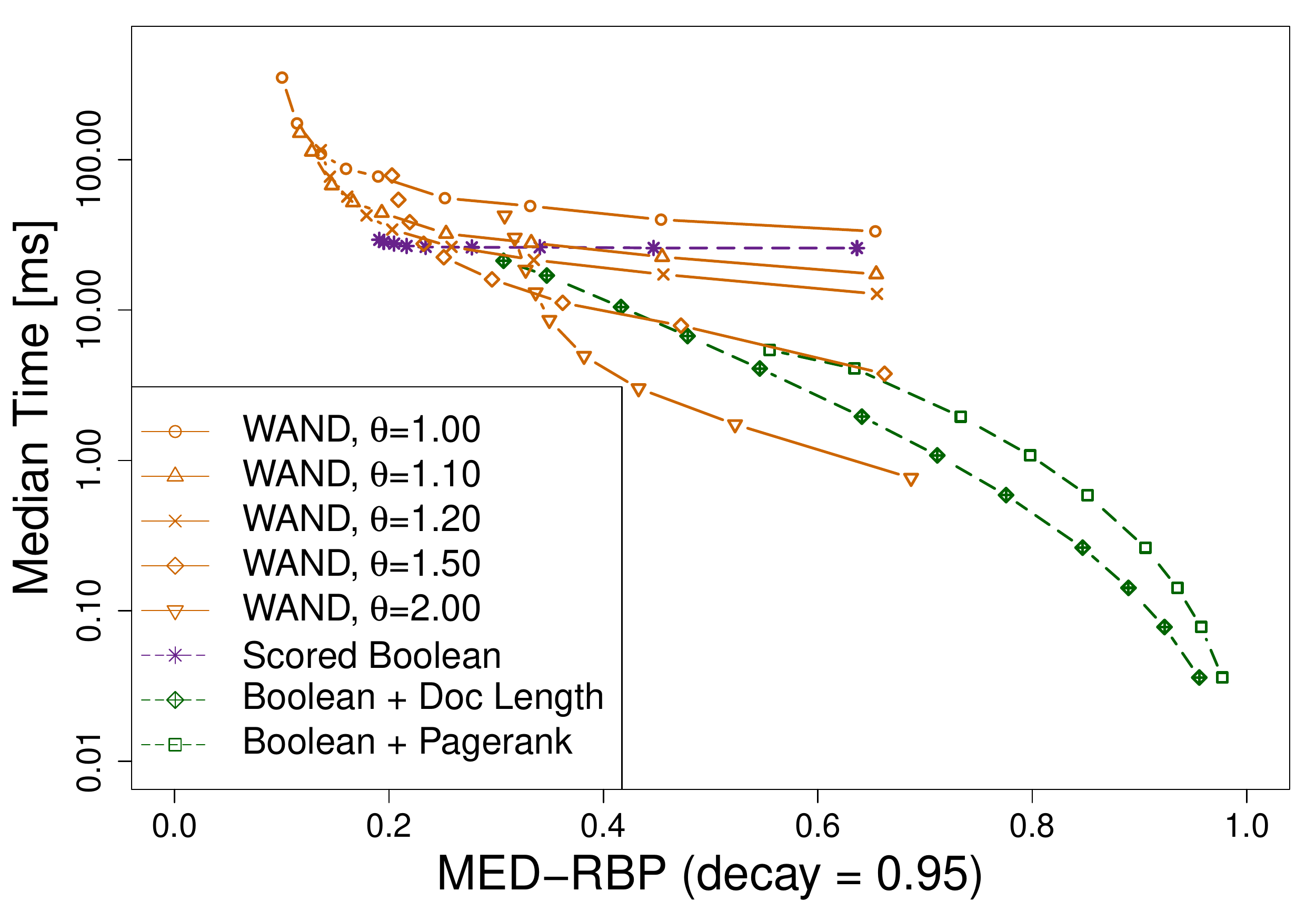}
\caption{Filter-stage effectiveness-efficiency trade-off curves, showing
the median query execution time as a function of mean $\med{RBP0.95}$
measured for a set of depths $k$, taken across $40{,}000$
queries on the stopped unpruned CW09B collection.
\label{fig-tradeoff}}
\end{figure}

\myparagraph{Filter-Stage Tradeoffs}
Figure~\ref{fig-tradeoff} depicts filter stage execution cost, now as
a function of $\med{RBP0.95}$, as $k$ is varied.
The suite of alternative methods describes a trade-off frontier that
defines a subset of techniques that are of possible interest.
That subset is dominated by aggressive WAND approaches, except when
$\metric{MED}$ is required to be very small.
The low fidelity of the static-score conjunctive Boolean methods
means that they do not contribute to the frontier except when query
processing must be very fast, in which case high $\metric{MED}$
scores must be tolerated.

\begin{figure}[t]
\centering
\showgraph{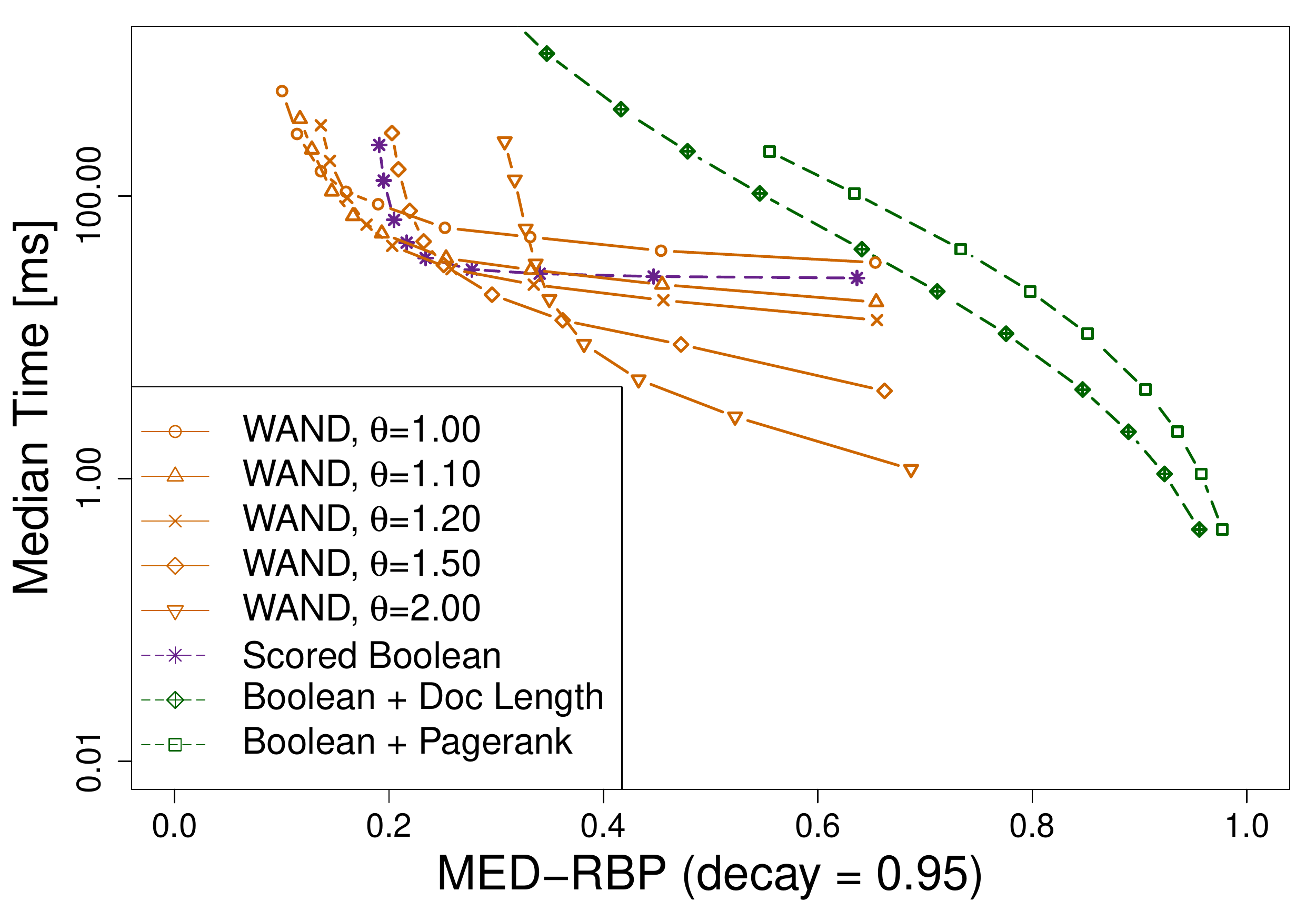}
\caption{Combined effectiveness-efficiency trade-off curves for
multi-stage retrieval, showing the median query execution time as a
function of mean $\med{RBP0.95}$ measured for a set of depths
$k$, taken across $40{,}000$ queries on the stopped unpruned CW09B
collection, and with the final ranking stage computation assumed to
require $0.02$ milliseconds per document.
\label{fig-money}}
\end{figure}

\myparagraph{Combining the Two Stages}
In a retrieval system the overall cost of generating a results page
and returning it to the user includes components other than the time
spent computing the filtered short-list of possible answers.
Most notably, the cost of performing the final ranking stage on the $k_1$
documents supplied by the filtering stage must be allowed for, and the
cost of creating answer snippets for the top $k_2$ documents identified
by the final ranking stage.
If the existence of a ``direct'' file is assumed, from which a
pre-generated set of features for any specified document can be
quickly retrieved, these costs are linear in $k_1$ and $k_2$
respectively; and with $k_1\gg{k_2}$, the cost of the final ranking
stage is the critical one.
{\citet{al13irj}} experimented with a variety of efficient index
representations and showed the average time per document to perform a
feature extraction in the ClueWeb09B collection was around $14$--$20$
$\mu$s.
The number of features required and the respective cost to retrieve
or calculate each largely depends on the final ranking stage algorithms 
used, but a $20$ $\mu$s per document penalty is a reasonable lower bound
for an efficient and effective final stage approach.

Figure~\ref{fig-money} is derived from Figure~\ref{fig-tradeoff}, and
is generating by adding an allowance for the final stage computation on
$k$ documents to each query's execution time, computed at a
(conservative) rate of $0.02$ milliseconds per document.
The Boolean conjunctive mechanisms are unattractive when viewed in
this light, and the aggressive WAND methods define the frontier for
best combinations of efficiency and effectiveness.
A very similar pattern of results was obtained with a final stage
scoring cost of $0.04$ milliseconds per document, indicating that the
overall relative costs are not particularly sensitive to the actual
value used.

Our work represents an extension and refinement of previous
measurements by {\citeauthor{al13irj}} {\citep{al13irj,al13sigir}}.
They also compared a range of filtering methods, including Boolean
conjunction and WAND-based disjunction.
Building on their results, we have used a faster Boolean computation
that allows simple early exit once $k$ documents have been
identified, and have also considered aggressive WAND techniques.
Moreover, we base our fidelity estimates on {\metric{MED}}
computations over large numbers of queries and do {\em not} require
relevance judgments; theirs are based on {\NDCG} scores that, as they
demonstrate, may be vulnerable to the uncertainties in measurement
associated with unjudged documents, and require relevance judgments
for all queries used in the final evaluation.


\section{Summary}
\label{sec-summary}

We have described a range of approaches that can be used for early
stage static and query-based filtering, with the purpose of selecting
a useful subset of a large collection to be ranked using a
final-stage reordering mechanism.
Our approach does not require the final stage output to be a proper
subset of the filtering stage output being evaluated, allowing
filters to be evaluated independently.
We have also described a methodology for quantifying the
effectiveness of a filtering stage that has the significant benefit
of not requiring relevance judgments.
Using this technique, and a large collection of queries and
documents, we have measured efficiency-effectiveness trade-offs in
multi-stage query systems.

In contrast to previous studies of efficiency-effectiveness
trade-offs, which were limited to smaller query sets, our methods are
applicable to sets containing thousands of queries.
As illustrated by the examples and figures of
Section~\ref{sec-tradeoffs}, our approach allows us to consider the
variance in query performance across these thousands of queries.
Identifying queries that perform particularly poorly provides
insights into the behavior of first-stage filters, potentially
leading to further improvements.
In particular, by training over large query sets, we may be able
select a early-stage filtering strategy, or even combinations of
filtering stages, for each query on an
individual basis by considering features derived from the index (for
example, the size of postings lists) and from the query itself (for
example, the number of query terms).
Note also that $\metric{MED}$ values can be monitored as queries are
processed, meaning that it might also be possible to develop a
feedback loop that reacts to mismatches as they are detected, and
switches to other (or deeper) filters.

Our experiments have made it clear that Boolean conjunction over all
query terms is not suitable as a filtering stage, even when coupled with
collection pre-ordering based on a static score such as PageRank or
document length.
That is, while $k$-prefixes of Boolean conjunctions can be computed
very quickly, that alone is insufficient to provide an interesting
multi-stage combination.
We also explored a range of WAND-based computations using Okapi
document scoring, including an aggressive WAND strategy and a scored
Boolean WAND approach, both of which do provide useful trade-offs.
By including the cost of a detailed final stage ranker, we were also
able to catalog the end-to-end cost of various combinations of filtering
stage and final ranking stage.
The results of this work demonstrate that the aggressive WAND
mechanism offer clear benefits for early stage filtering, since it
can be implemented to execute quickly, and provides a range of
combinations of $\theta$ and $k$ that can be balanced as required,
even on a query-by-query basis.


\myparagraph{Acknowledgments}
This work was supported by the Australian Research Council's
{\emph{Discovery Projects}} Scheme (DP140101587 and DP140103256).
Shane Culpepper is the recipient of an Australian Research Council
DECRA Research Fellowship (DE140100275).

\balance
\setlength{\bibsep}{3.5pt}
\bibliographystyle{abbrvnat}
\bibliography{strings-long,local} 


\end{document}